\documentclass[prl,twocolumn,superscriptaddress,showpacs,preprintnumbers,floatfix,amsmath,amssymb,nofootinbib,tightenlines]{revtex4}
\usepackage{graphicx}

\newcommand{\beqn}{\begin{eqnarray}}
\newcommand{\eeqn}{\end{eqnarray}}
\newcommand{\eq}[1]{(\ref{#1})}
\newcommand{\dd}{{\mathrm{d}}}
\newcommand{\Z}{{Z\!\!\!Z}}

\begin{document}

\preprint{ITEP-LAT/2005-26}
\preprint{LU-ITP 2005/025}
\preprint{HU-EP-05/79}
\preprint{UUITP-25/05}

\title{Phase structure of an Abelian two-Higgs model \\
and high temperature superconductors}

\author{M.\,N.~Chernodub}
\affiliation{
ITEP,
B.Cheremushkinskaya 25, RU-117259 Moscow, Russia}
\affiliation{Department of Theoretical Physics, Uppsala University,
P.O. Box 803, S-75108, Uppsala, Sweden}
\author{E.-M.~Ilgenfritz}
\affiliation{Institut f\"ur Physik, Humboldt-Universit\"at zu Berlin,
Newtonstr. 15, D-12489 Berlin, Germany}
\author{A.~Schiller}\affiliation{Institut f\"ur Theoretische Physik,
Universit\"at Leipzig, D-04109 Leipzig, Germany}

\begin{abstract}
We study the phase structure of
a three dimensional Abelian Higgs model with singly- and doubly-charged
scalar fields coupled to a compact Abelian gauge field. The model is
pretending to describe systems of strongly correlated electrons
such as high-$T_c$ superconductivity in overdoped regime and exotic phases
supporting excitations with fractionalized quantum numbers.
We identify the Fermi liquid, the spin gap, the superconductor and the strange
metallic phases in which
densities and properties of holon and spinon vortices and monopoles
are explored. The phase diagram in the $3D$ coupling space is predicted.
We show that at sufficiently strong gauge coupling
the spinon-pair and holon condensation transitions merge together
and become, unexpectedly,
first order.
\end{abstract}

\pacs{74.10.+v,71.10.Hf,11.15.Ha,74.90.+n}

\maketitle

The physics of high-$T_c$ superconductivity~\cite{ref:review:Dagotto,ref:review}
is not yet completely understood. However, certain outstanding features
are generally recognized. At normal temperatures, all known high-$T_c$
superconductors are ceramic materials characterized by a poor conductivity.
At low temperatures the clean material is a Mott insulator. Doping it
with impurities, the insulator may become a superconductor at
low enough temperatures. The physics of high-$T_c$ superconductivity
is essentially a $2D$ phenomenon since charge
carriers -- which are electrons or holes provided by the dopants --
are confined to the CuO${}_2$ planes
typical for all high-$T_c$ superconductors.

A popular approach to high-$T_c$ superconductivity is provided by the $t$-$J$
Hamiltonian~\cite{ref:Anderson} describing hopping holes and localized spins
in a plane:
\beqn
  H_{tJ} = - t \!\! \sum\limits_{<ij>,\sigma} \!\!
  c_{i\sigma}^\dagger P_{ij,-\sigma} c_{j\sigma}
  + J \!\sum\limits_{<ij>} ({\vec S}_i {\vec S}_j - \frac{1}{4} n_i n_j)
  \,.
  \label{eq:tJ}
\eeqn
The first term describes
electrons moving without changing spin $\sigma$.
Double occupancy is explicitly forbidden by the presence of the projectors
$P_{ij,\sigma} = (1 - n_{i,\sigma})(1 - n_{j,\sigma})$.
The second term specifies the
anti-ferromagnetic Heisenberg coupling between
spins located at the copper sites. Here
$\vec S_i=(1/2) \sum_{\sigma\sigma'}c_{i\sigma}^\dagger {\vec \tau}_{\sigma\sigma'}  c_{i\sigma'}$
is the spin operator, $c^{\dagger}_{i\sigma}$, $c_{i\sigma}$
are the hole (or electron) creation and annihilation operators,
$n_{i,\sigma} = c_{i\sigma}^\dagger c_{i\sigma}$ denotes the occupation number,
and $n_i= n_{i,\uparrow} + n_{i,\downarrow}$.

The $t$-$J$ model~\eq{eq:tJ} is often treated in the slave-boson
technique~\cite{ref:slave-boson}, which splits the spin and charge degrees of
freedom of the electrons.  The electron creation operators are decomposed
as~\cite{ref:slave-boson,ref:Baskaran:Solid:State}
$c^\dagger_{i\sigma} = f^\dagger_{i\sigma} b_i$, where $f_{i\sigma}$ is a
spin-particle (``spinon'') operator and $b_{i}$ is a charge-particle
(``holon'') operator. In order to forbid double occupancy of sites one
imposes the constraint
$f^\dagger_{i\uparrow} f_{i\uparrow} + f^\dagger_{i\downarrow} f_{i\downarrow} + b^\dagger_{i} b_{i} = 1$
on the physical states of the system.
The constraint drastically simplifies~\cite{ref:Baskaran:Solid:State}
the treatment of the $t$-$J$ Hamiltonian~\eq{eq:tJ}.

In addition to the ordinary electromagnetic (external) gauge symmetry,
the spin-charge separation naturally introduces an (internal) compact $U(1)$ gauge
freedom,
\beqn
  U(1)_{\mathrm{int}}:\ \, \
  c_{i\sigma} \to c_{i\sigma},\ \,
  f_{i\sigma} \to e^{i \alpha_i} f_{i\sigma}, \ \,
  b_{i} \to e^{i \alpha_i} b_{i\sigma}
  \label{eq:gauge:internal}
\eeqn
which plays an essential role~\cite{ref:Baskaran} in understanding the physics
of strongly correlated electrons.
Besides condensed matter physics, the spin-charge separation idea is also
applied to strongly interacting gluons in QCD~\cite{ref:slave-boson-further}.

The emerging effective theory of superconductivity can further be
simplified and reformulated in terms of lattice gauge
models~\cite{ref:Baskaran,ref:LeeNagaosa:characterization,NagaosaLee,ref:Ichinose:Matsui}.
Thus, the $t$-$J$ model~\eq{eq:tJ} is related to a compact Abelian
gauge model with internal symmetry~\eq{eq:gauge:internal}, which
couples holons and spinons.
As in usual BCS superconductivity, under appropriate conditions
the spinons couple and form
bosonic quasiparticles. In a mean field theory one can define fields
which behave under the gauge transformations~\eq{eq:gauge:internal} as:
\beqn
  \chi_{ij} & = & \sum\nolimits_\sigma \langle f^\dagger_{i\sigma} f_{j\sigma} \rangle
  \to \chi_{ij} \cdot
  {\mathrm e}^{- i (\alpha_i - \alpha_j)}\,, \\
  \Delta_{ij} & = & \langle  f_{i\uparrow}f_{j\downarrow}
                       - f_{i\downarrow}f_{j\uparrow}  \rangle
  \to \Delta_{ij} \cdot {\mathrm e}^{i (\alpha_i + \alpha_j)}\,.
\eeqn
The phase of the field $\chi$ is the compact
$U(1)$ gauge field, $\theta_{ij} \equiv \arg \chi_{ij} \to
\theta_{ij} + {(\dd \alpha)}_{ij}$ with ${(\dd \alpha)}_{ij} = \alpha_j - \alpha_i$,
and the radial part, $\chi = | \langle \chi_{ij} \rangle |$,
is the so-called ``resonating valence bond'' (RVB) coupling.
The doubly-charged spinon-pair field $\Delta$ is an analog of
a Cooper pair.

At high temperature the RVB coupling is vanishing, $\chi=0$, and
the system is in the Mott insulator (or ``poor metallic'') phase.
With decreasing temperature $\chi$ becomes non-zero,
eventually enabling the formation of a spinon-pair condensate
$\Delta = |\langle\Delta_{ij}\rangle|$ and/or of a  holon condensate
$b = \langle b_i\rangle$~\cite{ref:Baskaran:Solid:State}.
Therefore, four phases~\cite{ref:LeeNagaosa:characterization,NagaosaLee,ref:Ichinose:Matsui}
may emerge: the Fermi liquid (FL) phase with $b \ne 0$, $\Delta=0$,
the spin gap (SG) phase with $b =0$, $\Delta\ne 0$,
the superconductor (SC) phase with $b \ne 0$, $\Delta\ne 0$, and the
strange metallic (SM) phase with $b =0$, $\Delta= 0$.

A compact Abelian two-Higgs model (cA2HM) in {\it three} dimensions
with a $U(1)$ gauge link field $\theta$, a
singly-charged holon field $\Phi_1$, and a
doubly-charged spinon-pair field $\Phi_2$
is a feasible phenomenological model to describe
the phase structure of the {\it two}-dimensional
spin-charge separated quantum
system~\cite{ref:LeeNagaosa:characterization,NagaosaLee}.
This model suffers from several limitations~\cite{ref:review}
to describe high-$T_c$ materials in the underdoped regime, since
the model includes only Gaussian fluctuations about the mean field
theory of the $t$-$J$ model and does not reflect the $SU(2)$ particle-hole
symmetry at half-filling. However, it seems plausible that the cA2HM
can be applied in the less-studied overdoped regime
(called sometimes the SM region~\cite{ref:review,ref:strange:metal})
where the particle-hole symmetry is explicitly broken.

Since high-$T_c$ materials are type-2 superconductors, we
restrict ourselves to the London limit in which the radial parts of both
Higgs fields $\Phi_Q = |\Phi_Q| \, e^{i \varphi_Q}$
are frozen, $|\Phi_{1,2}| = {\mathrm{const}}$.
The action of the model is defined as
\beqn
S_{\mathrm{A2HM}} = -\beta \sum\limits_P \cos\theta_P
                    - \sum_{Q=1}^2 \kappa_Q
                      \sum\limits_l \cos({\mathrm{d}} \varphi_Q + Q \theta)_l ,
  \label{eq:cA2HM:action}
\eeqn
where
$\theta_P$ is the standard lattice plaquette.
The model~\eq{eq:cA2HM:action}
obeys a lattice version of the $U(1)$ internal gauge
symmetry~\eq{eq:gauge:internal}:
$\theta \to \theta + {\mathrm{d}} \alpha$, $\varphi_Q \to \varphi_Q - Q \alpha$.
It describes the hole (electron) ``constituents'' by the dynamical holon  $\varphi_1$ and
spinon $\varphi_2$ phases, which strongly interact via the dynamical gauge
field $\theta$. The inverse gauge coupling $\beta$ in Eq.~\eq{eq:cA2HM:action},
$\beta = \chi_f + \chi_b$, is given by the diamagnetic susceptibilities
$\chi_f$ of the spinon and $\chi_b$ of the holon fields.
The hopping parameters $\kappa_Q$ are connected to
the doping concentration $x$ and the couplings $t$ and $J$ of (\ref{eq:tJ})
as follows:
$\kappa_1 \propto t \cdot x$ and $\kappa_2 \propto J$~\cite{ref:review}.

Multi-Higgs models similar to Eq.~\eq{eq:cA2HM:action} appear not only
for theories of high-$T_c$ superconductivity but also in
models for metallic hydrogen~\cite{ref:metallic}
(using non-compact gauge fields and $Q_1 = - Q_2 = 2$)
and for multi-band superconductors~\cite{ref:two:component}.

The model~\eq{eq:cA2HM:action} contains three kinds of topological defects:
a monopole and two types of vortices, referred to as the holon and the
spinon vortex~\cite{ref:review,ref:LeeNagaosa:characterization}.
The monopole has magnetic charge $2 \pi \hbar$ while the holon (spinon) vortex
carries magnetic flux quanta $2 \pi \hbar$ ($\pi \hbar$) of the
gauge field $\theta$. Since the magnetic flux is conserved,
one monopole is simultaneously a source of one holon vortex and two
spi\-non vortices. The defects can be best studied in the dilute state
characterizing the superconducting phase of the model with nonvanishing condensates.
In this phase, the core of the holon (spinon) vortex is in the FL (SG) phase.
The monopole core is in the SM phase where both
condensates are zero.

\begin{figure}[!htb]
\vspace{1mm}
\begin{tabular}{cc}
  \includegraphics[scale=0.30,clip=true]{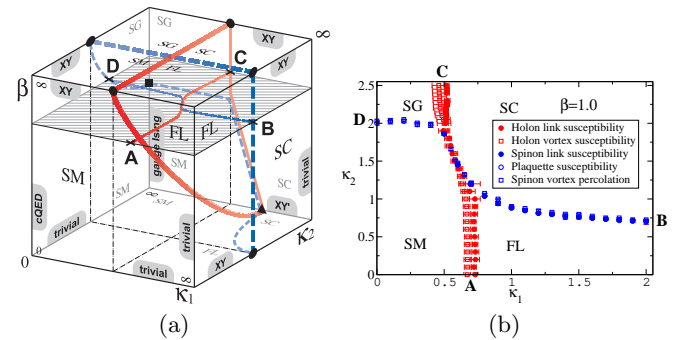} &
  \includegraphics[scale=0.17,clip=true]{beta10crit_paper.eps} \\
(a) & (b)
\end{tabular}
\caption{(a) The qualitative $3D$ phase diagram and (b) its numerically obtained
$2D$ cross-section at $\beta=1$.}
\label{fig:phase:3D}
\end{figure}

The structure of the three-dimensional phase diagram of the
model~\eq{eq:cA2HM:action} is quite complicated. However, the
faces and edges of the $3D$ phase ``cube''
({\it i.e.} the limiting cases of vanishing or large couplings
$\beta$ and $\kappa_{1,2}$), respectively, can be related to various well-known
and sometimes non-trivial systems. These limiting cases are shown schematically
on the faces of the $3D$ phase cube presented in Fig.~\ref{fig:phase:3D}(a)
and are discussed in detail below.
The interior of the cube in Fig.~\ref{fig:phase:3D}(a) is our qualitative
prediction of the phase structure of the cA2HM. The shaded $(\kappa_1,\kappa_2)$
plane should be compared with Fig.~\ref{fig:phase:3D}(b) which represents the result
of our numerical investigation for $\beta=1.0$.
The numerical result is consistent with the prediction and
will be presented later.

On the $\kappa_2=0$ face the compact Abelian $Q=1$ Higgs model
(cAHM${}_{Q=1}$) emerges. There the holon condensate $b$ and the holon vortices
are controlled by the compact $U(1)$ gauge field.
The phase of the spinon-pair field is disordered which implies
condensation of the spinon vortices and, therefore, vanishing $\Delta$.
In the ($\beta$,$\kappa_1$) plane two
phases~\cite{ref:fradkin,Osterwalder:1977pc,ref:einhorn,ref:AHM1} exist:
(i) the broken/Higgs phase with $b \neq 0$  at
large $\beta$/large $\kappa_1$ corresponding to the FL phase,
and (ii) the confining/symmetric phase with $b\approx 0$ at
small $\beta$/small $\kappa_1$. The confining phase is the SM phase with
some properties of the FL phase.
The holon condensate is nowhere exactly vanishing
because the phases are connected analytically.
The holon vortices are almost massless and therefore dense in the SM
phase, and they become heavy and dilute in the FL phase.
At the $\kappa_1 =0$ edge the model reduces to the compact
$U(1)$ theory which is confining at any $\beta$ due to the monopole
plasma~\cite{ref:Polyakov}. The plasma destroys the holon condensate, $b=0$.
At the edge $\beta \to \infty$ the model becomes the $3D$ $XY$ model
with a second order transition~\cite{ref:XYphase} at
$\kappa_{1,c} = \kappa^{XY}_c \approx 0.45420\dots$
Since away from this limit a relevant global symmetry is absent,
no singularities in the thermodynamic quantities are expected.
With decreasing $\beta$, the SM and FL phases, being analytically connected,
can be thought to be separated by a Kert\'esz line~\cite{Wenzel:2005nd},
across which the holon condensate is a smooth quantity.

On the face $\kappa_2 \to \infty$ the model (\ref{eq:cA2HM:action}) reduces
to a $3D$ $XY$-$\Z_2$ model with the action
\beqn
S_{XY-\Z_2} = -\beta \sum\nolimits_P \sigma_P -
\kappa_1 \sum\nolimits_l \sigma_l \cos(d\varphi)_l\,.
\label{eq:XYZ2:action}
\eeqn
This model describes a $XY$-type matter field $\varphi = \varphi_1 - \varphi_2/2$
which interacts via exchange of a $\Z_2$ gauge field $\sigma_l = \pm 1$.
The model -- studied numerically in Ref.~\cite{ref:Z2XY} --
contains three phases which, according to the adopted classification, are:
(i)   the SM phase at small $\beta$/small $\kappa_1$,
(ii)  the SG phase at large $\beta$/small $\kappa_1$ and
(iii) the SC phase at large $\beta$/large $\kappa_1$.

For not too large $\kappa_1$ the SM and SG phases are separated by a second
order phase transition of the Ising universality class
across a line $\beta_c(\kappa_1)$
with the $3D$ gauge Ising (gI) model limit~\cite{ref:Z2gauge}:
$\beta_c(0) =\beta^{\mathrm{gI}}_c \approx 0.7613\dots$
At large $\beta$ the SG phase and SC phases are separated by a second order phase
transition in the $3D$ $XY$ universality class,
$\lim_{\beta \to \infty} \kappa_{1,c}(\beta) = \kappa^{XY}_c$.
Inside the phase diagram these transition lines merge and continue as a single
second order transition in the $XY$ universality class along a line
$\kappa_{1,c}(\beta)$ which ends at the $\beta=0$ edge in
$\kappa_{1,c}(0) = \kappa^{XY'}_c \approx 1.6$.
Here the model is reduced to a (modified) $XY$ model with
the action $S_{XY'} = - \sum_l \log \cosh [\kappa_1 \cos (\dd \varphi)_l]$.

The model (\ref{eq:XYZ2:action})
has exotic phases predicted for strongly correlated
electron systems~\cite{ref:fractionalized} and it is characterized by so-called
``visons'' -- fractionally charged excitations related to the presence
of $\Z_2$ degrees of freedom. A vison coincides with the spinon vortex at
$\kappa_2 \to \infty$, while the holon vortex
turns into a $XY$ vortex. The $XY$ phase angle $\varphi$ becomes
the phase of the so-called ``chargon'' particle.
The SG phase corresponds to the fractionalized phase where visons are absent
and chargons are free particles.
In the SM phase the visons are condensed, and chargons are confined.
The SC phase corresponds to a superfluid state where both visons and $XY$
vortices are dilute and chargons are free.

On the $\kappa_1 = 0$ face the holon condensate vanishes, $b = 0$, and
the spinon-pair condensate $\Delta$ is described by the cAHM${}_{Q=2}$ model.
The phase diagram in the ($\beta$,$\kappa_2$) plane can be deduced from
results of
Ref.~\cite{ref:fradkin,Smiseth:2003bk}:
The SM phase is located at small $\beta$/small $\kappa_2$,
while the SG phase is residing in the large $\beta$/large $\kappa_2$ corner.
The phases are separated by a second order phase transition line which ends at
$\beta_c = \beta^{\mathrm{gI}}_c$ in the large $\kappa_2$ limit and at
$\kappa_{2,c} = \kappa^{XY}_c$ in the large $\beta$ limit.

On the $\kappa_1 \to \infty$ face the model (\ref{eq:cA2HM:action})
reduces to a $XY$ model,
$S_{XY} = - \kappa_2 \sum\nolimits{_l} \cos(\dd \varphi)_l$,
$\varphi= 2 \varphi_1 +\varphi_2$, which controls the spinon condensate.
Due to the constraint $\theta = \dd \varphi_1 + 2 \pi l$, $l \in \Z$,
the holon vortices are suppressed and therefore $b \neq 0$ in the whole
($\beta$,$\kappa_2$) plane. The phase diagram is divided by
a straight line of a second
order $XY$-like transition parallel to the $\beta$ axis
at $\kappa_{2,c}(\beta) = \kappa^{XY}_c$.
This line separates the FL phase (with condensed spinon vortices and $\Delta=0$)
at $\kappa_2 < \kappa^{XY}_c$ from the SC phase (with suppressed spinon vortices
and $\Delta\ne0$).

On the $\beta=0$ face we obtain a two-Higgs system interacting
ultra-locally via a non-propagating gauge field.
After integrating out the gauge field in the Villain
representation~\cite{ref:Villain}, one gets a $XY$ model with the effective
$XY$-coupling,
$\kappa^V_{\mathrm{eff}} = \kappa^V_1 \kappa^V_2/ (\kappa^V_1 + 4 \kappa^V_1)$
where the superscript $V$ indicates the Villain coupling. The phase diagram
in the ($\kappa_1$,$\kappa_2$) plane contains two phases: the SC phase
with non-zero condensates $\Delta$ and $b$
in the large $\kappa_1$/large $\kappa_2$ corner
and a SM-FL phase in the remaining part of the phase diagram.
The phases are separated by a second order $XY$-type phase transition which starts
at $\kappa_{1,c} = \kappa^{XY'}_c$ at the $\kappa_2 \to \infty$ edge
and ends at $\kappa_{2,c} = \kappa^{XY}_c$ for $\kappa_1 \to \infty$.
At these two edges
the model is reduced to a modified and a usual $XY$ model, respectively.
The SM-FL phase is actually the FL phase at large $\kappa_1$,
where the holon condensate is non-vanishing, and the $SM$ phase
at large $\kappa_2$, where the holon condensate tends to zero. Those phases are
analytically connected, similar to the $\kappa_2 = 0$ face considered above.

On the $\beta \to \infty$ face the system reduces to two decoupled $XY$ models
describing the holon and spinon superfluidities.
The phase diagram in $(\kappa_1,\kappa_2)$ includes all SM, SG, FL, SC phases as
shown on the upper face of Fig.~\ref{fig:phase:3D}(a).
The predicted $3D$ phase diagram implies, in particular, that at sufficiently
strong gauge coupling, $\beta < \beta^{\mathrm{gI}}_c$, the SG phase ceases
to exist.

Performing Monte Carlo simulations we have investigated the phase diagram of the
cA2HM. We made an exploratory
first study on a $16^3$ lattice,
choosing three values of the gauge couplings,
$\beta=1.0$, $1.5$ and $2.0$, on a
dense grid of points spanning the
($\kappa_{1}$,$\kappa_2$) hopping parameter plane over the range
$0 < \kappa_{1,2} \leqslant 2.0,2.5$.
In all figures we show the results for the strong coupling $\beta=1.0$ only.

\begin{figure}[!htb]
  \begin{tabular}{cc}
\multicolumn{2}{c}{\includegraphics[scale=0.25,clip=true]{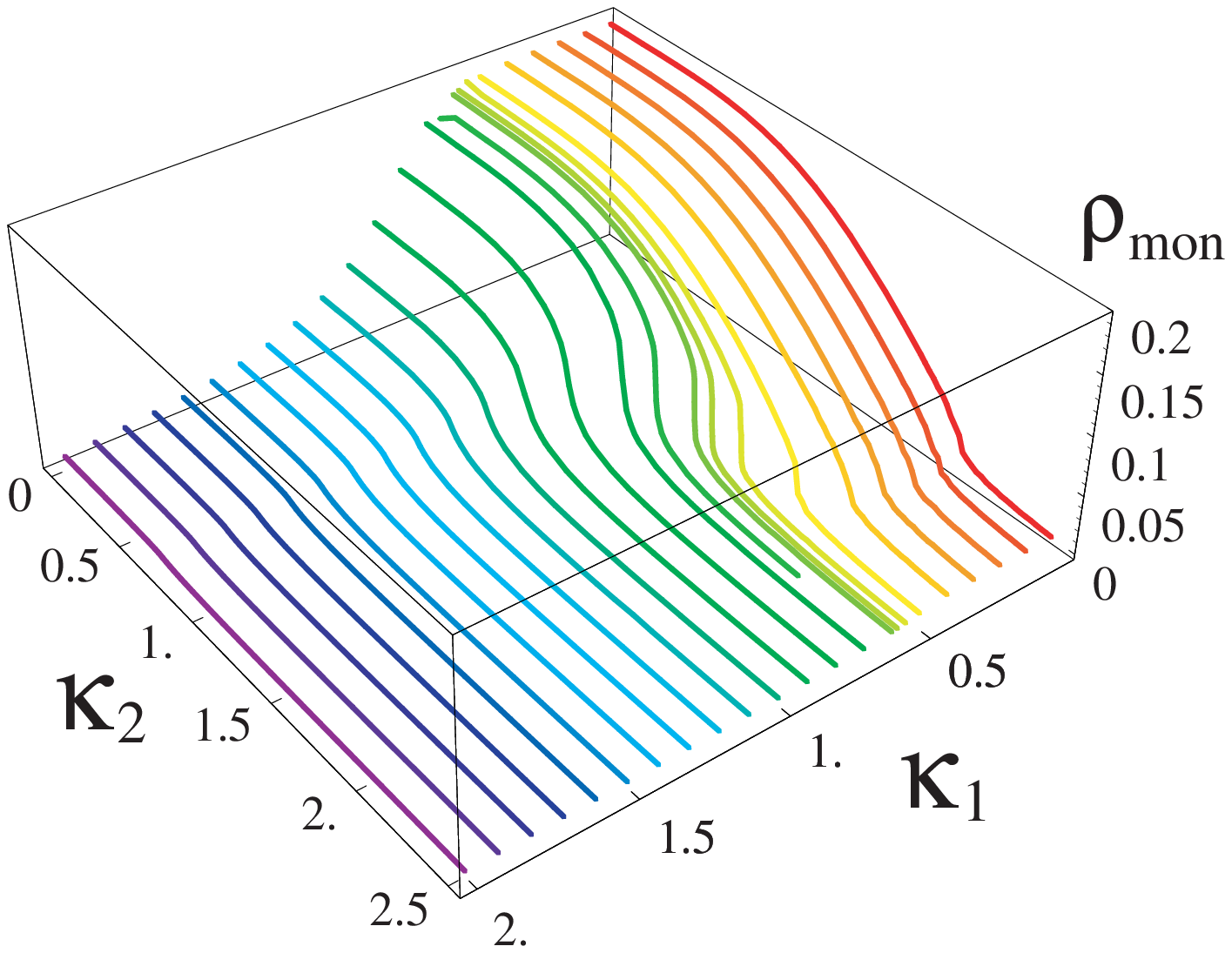}}\\[-5mm]
    \includegraphics[scale=0.25,clip=true]{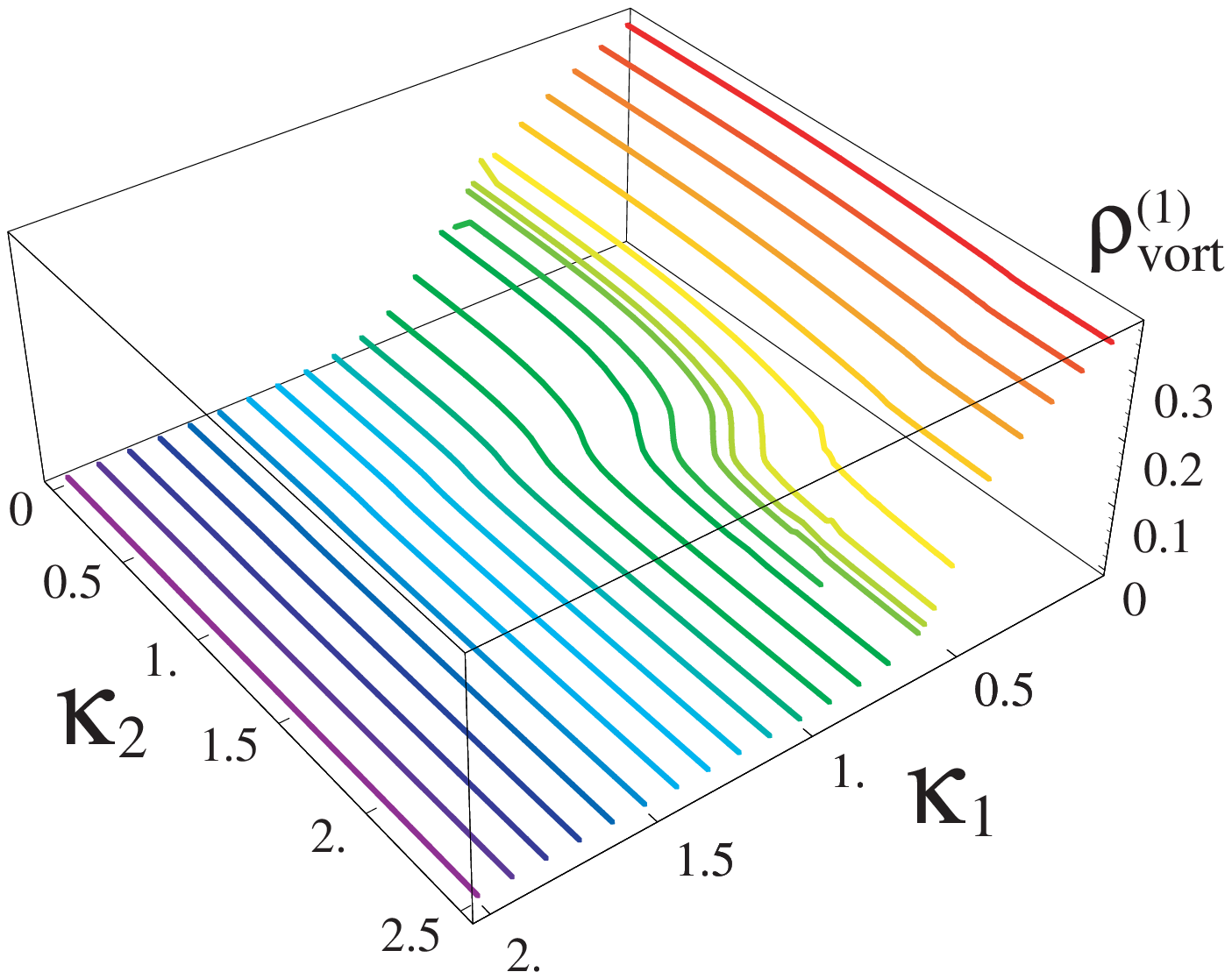} &
    \includegraphics[scale=0.25,clip=true]{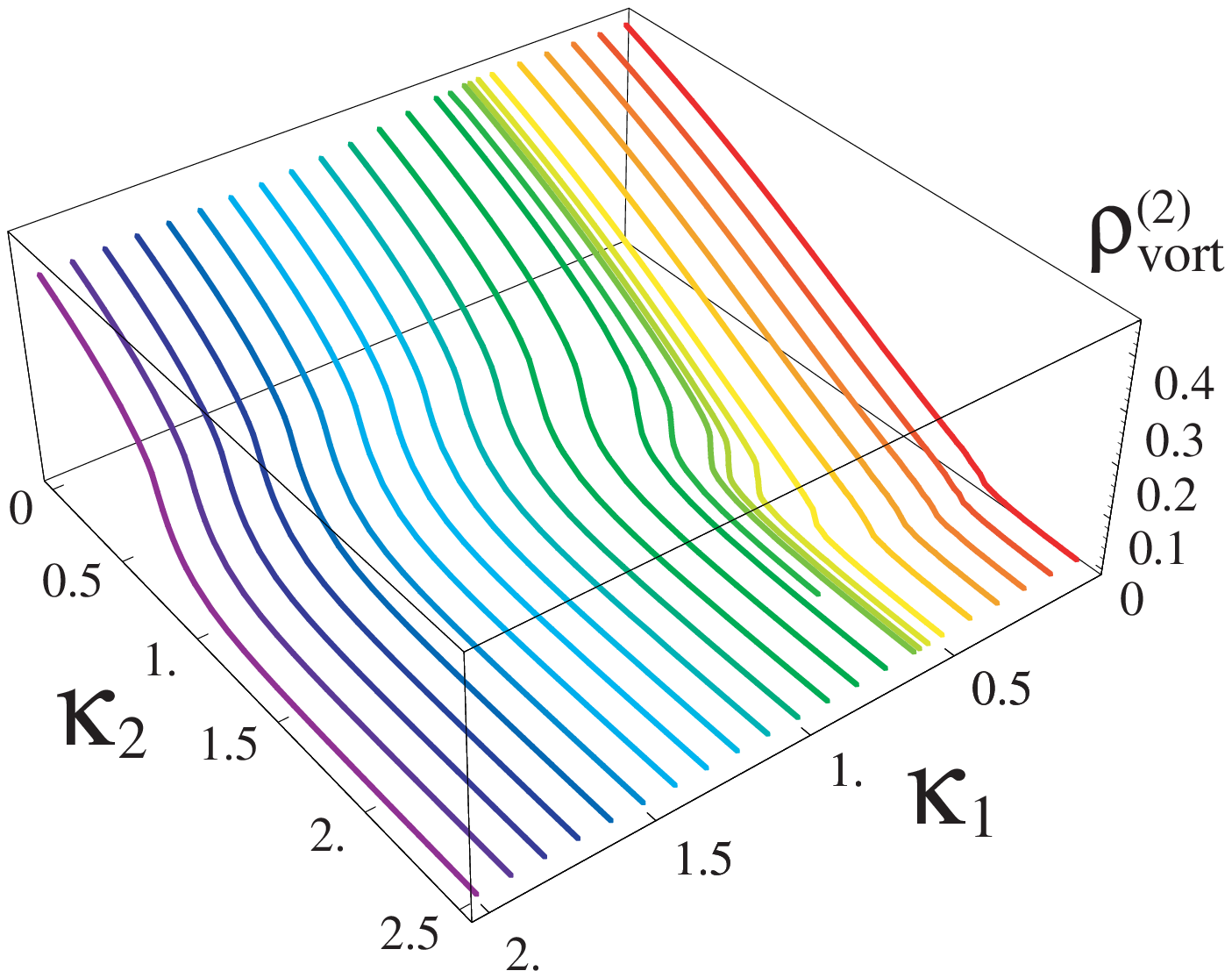} \\[-10mm]
    \includegraphics[scale=0.25,clip=true]{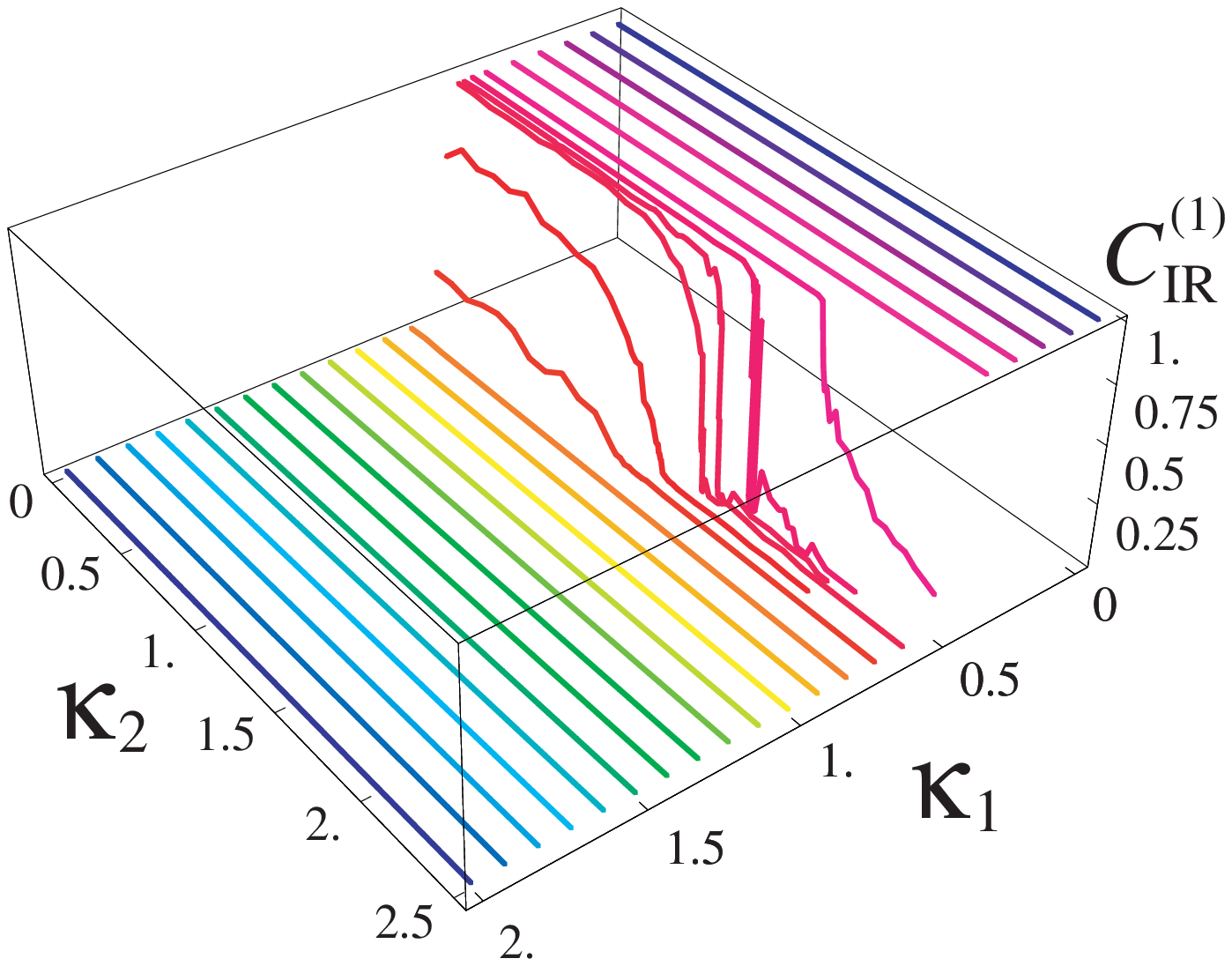} &
    \includegraphics[scale=0.25,clip=true]{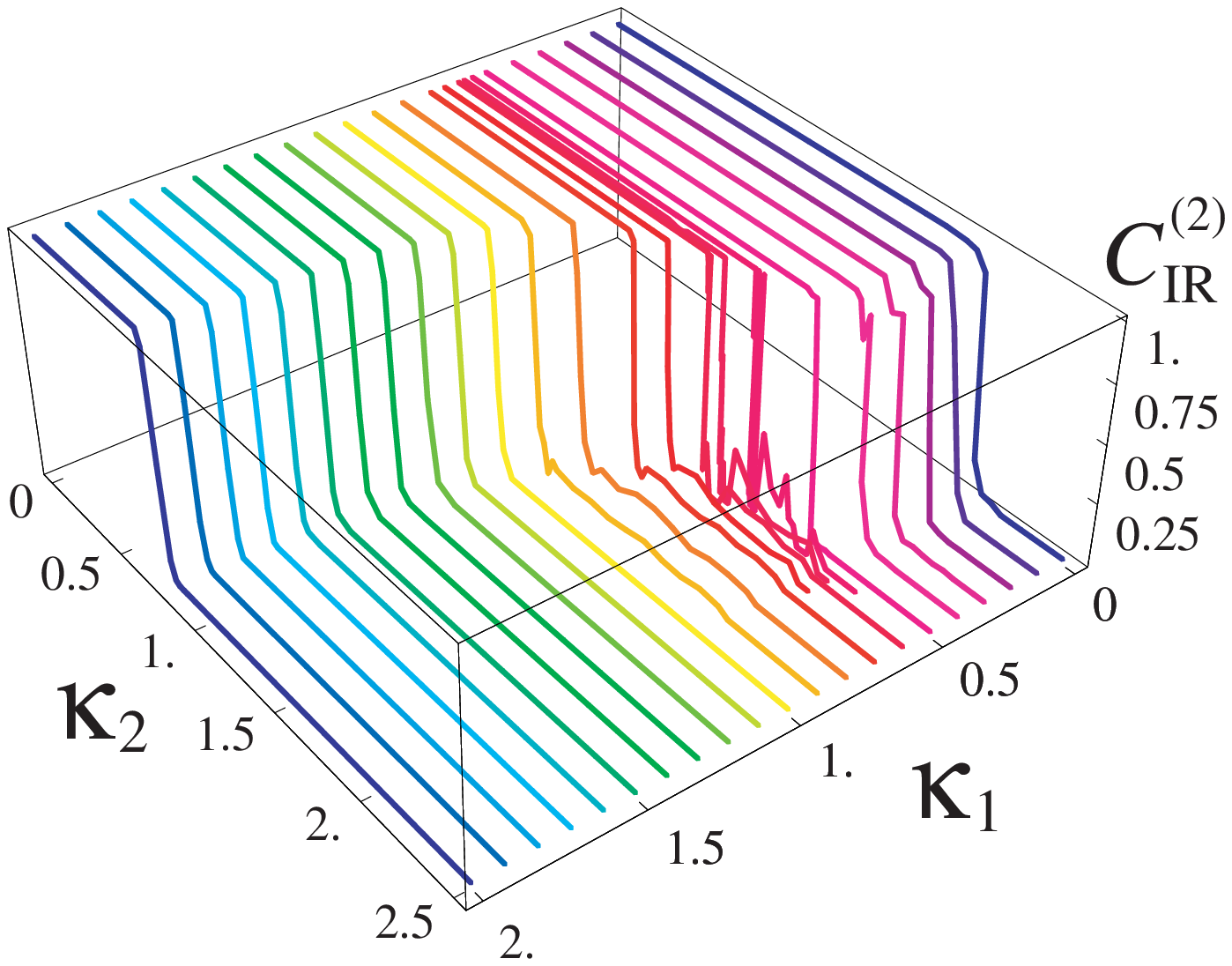} \\
   holon vortex& spinon vortex
  \end{tabular}
  \caption{The monopole density (top), the vortex densities (middle), and
  the vortex connectivities (bottom). Left: holon vortices; right: spinon
  vortices.
   }
  \label{fig:density:IR}
\end{figure}

The density of the monopoles, $\rho_{\mathrm{mon}}$, is
plotted in the upper panel of Fig.~\ref{fig:density:IR}
over the ($\kappa_1$,$\kappa_2$) plane.
With increasing hopping parameters $\kappa_1$ or $\kappa_2$
the monopole density gets suppressed.
As shown in the middle panel of Fig.~\ref{fig:density:IR}
the density of the holon vortices $\rho^{(1)}_{\mathrm{vort}}$
(spinon vortices $\rho^{(2)}_{\mathrm{vort}}$)
significantly drops down with increasing $\kappa_1$ ($\kappa_2$).

The connectivity $C^{(1,2)}_{\mathrm{IR}} = C^{(1,2)}(R_{\mathrm{max}})$ of the vortex clusters
is derived from the cluster correlation functions
\beqn
C^{(1,2)}(|{\vec R}|) = \frac{ \langle \sum_{{\vec x},{\vec y}} \sum_{c}
\Theta_{c}({\vec x})
\Theta_{c}({\vec y}) \delta({\vec R}-{\vec x}+{\vec y}) \rangle }
{\sum_{{\vec x},{\vec y}} \delta({\vec R}-{\vec x}+{\vec y})}
\eeqn
($\Theta_{c}({\vec x})=1$ or 0 is the characteristic function of the vortex cluster
$c$, $R_{\mathrm{max}}$ the maximal distance on the periodic lattice)
gives a clear view of the phase diagram of the
model~\cite{ref:AHM1,ref:AHM2,Wenzel:2005nd}.
Any Higgs condensate suppresses the proliferation of the corresponding
vortices: they are prevented to percolate over infinitely long distances.
We show the connectivities $C^{(1,2)}_{\mathrm{IR}}$ for the spinon (holon)
vortex clusters in the lower panel of Fig.~\ref{fig:density:IR}.
The corresponding phase diagram is shown in Fig.~\ref{fig:phase:3D}(b).
It agrees with the cross-section through the $3D$ phase cube marked by
the shaded plane in Fig.~\ref{fig:phase:3D}(a).
We plotted letters $A$, $B$, $C$ and $D$ in Figs.~\ref{fig:phase:3D}(a) and (b)
to identify the same points.

To clarify the nature of the phase transitions, we studied the volume dependence
of the average plaquette and both link contributions to the
action~(\ref{eq:cA2HM:action}) as well as their susceptibilities, respectively,
in different regions of the phase diagram at $\beta=1.0$.
\begin{figure}[!htb]
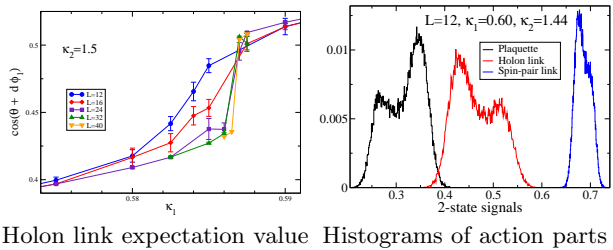

  \begin{tabular}{cc}
    \includegraphics[scale=0.16,clip=true]{first_holonlink_k1_150.eps} &
    \includegraphics[scale=0.16,clip=true]{hist_proc.eps} \\
    Holon link expectation value & Histograms of action parts
  \end{tabular}
  \caption{Signatures for the first order transition.
  }
  \label{fig:thermodyn1}
\end{figure}
Fig.~\ref{fig:thermodyn1} (left) clearly exemplifies the
first order nature of the transition (observed in all parts of action)
by the jump developing in the holon link
vs. $\kappa_1$ at
fixed $\kappa_2$ in the crossing region of the transition lines.
Already at lattice size $32^3$ we do not observe tunnelings for the
selected $\kappa_1$ values within $ 5 \times 10^5$ iterations. In
Fig.~\ref{fig:thermodyn1} (right) we present typical two-state
signals (for a $12^3$ lattice) of all three terms in the action
at $(\kappa_1,\kappa_2)$ near the transition.
The two-state signal with respect to the plaquette
and the holon link becomes very weak when one goes to smaller $\kappa_1$ along the
(except for the direct transition between the SM and the SC phases) horizontal
dark-dotted (blue) line.

It is known that for $\kappa_1 \to 0$ the transition {\it vs.} $\kappa_2$
is of second order.
Similarly, for large $\kappa_1$ the transition is most likely of second order,
again in the $XY$ universality class. We found that for the largest
volumes $40^3-48^3$ under study the
increase of the spinon link susceptibility at $\kappa_1=2.0$
stops as function of the lattice size
as expected for the $XY$ model at $\kappa_1\to \infty$~\cite{Campostrini:2000iw}.
Selecting two $\kappa_2$ values outside the crossing regime,
where one of the transition lines
(the light-dotted [red] one) runs vertical,
we observe that there is no thermodynamic
transition {\it vs.} $\kappa_1$ for
the smaller $\kappa_2$, in agreement with what could be anticipated from
the limit $\kappa_2 \to 0$.

Summarizing, we predicted the qualitative $3D$ phase diagram of the
cA2HM proposed as an effective model for high-$T_c$ cuprates in the
overdoped regime. At strong gauge coupling
(however $\beta > \beta^{\mathrm{gI}}_c$)
we found two transitions
associated with the pattern of vortex percolation
and identified the Fermi liquid, spin gap, superconductor and strange metallic
phases in agreement with the proposed phase diagram.
The percolation transitions are accompanied with thermodynamic
phase transitions except for small holon hopping parameter.
First hints for a changing order along the thermodynamic transition
lines are found. In the region where the two transition lines merge,
a direct phase transition between the SC and SM phases
exists and is found to be first order.

\begin{acknowledgments}
E.-M.I. is supported by the DFG
Forschergruppe 465 ``Gitter-Hadronen-Ph\"anomenologie''.
M.N.Ch. is supported by a STINT Institutional grant IG2004-2 025
and by the grants RFBR 04-02-16079, 05-02-16306, DFG 436 RUS
113/739/0, and MK-4019.2004.2.
M.N.Ch. is thankful to members of Department of Theoretical Physics of
Uppsala University for kind hospitality and stimulating environment.
\end{acknowledgments}

\end{document}